\documentclass[aps,prb,twocolumn,showpacs,superscriptaddress,amsmath]{revtex4}
\usepackage{graphicx}

\begin{document}

\title{Molecular and dissociative absorption of  multiple  
hydrogens on transition metal decorated C$_{60}$}

\author{T. Yildirim}
\email{taner@nist.gov}
\affiliation{NIST Center for Neutron Research, National
Institute of Standards and Technology, Gaithersburg, MD 20899}

\author{Jorge \'I\~niguez}

\affiliation{Institut de Ci\'encia de Materials de Barcelona (CSIC),
Campus de la UAB, 08193 Bellaterra, Barcelona, Spain}

\author{S. Ciraci}

\affiliation{Physics Department, Bilkent University, 
06800 Bilkent, Ankara, Turkey}

\begin{abstract}
Recently we have predicted [Phys. Rev. Lett. May 2005] that Ti-decorated
carbon nanotubes can absorb up to 8-wt\% hydrogen at ambient conditions.
 Here  we show that  similar phenomena 
occurs in light transition-metal decorated C$_{60}$. 
%We consider
%four possible absorption sites on a C$_{60}$ molecule; namely
%the hexagonal (H) and pentagonal (P) hollow sites, top of double (D) and
%single (S) bonds. 
While Sc and Ti prefers the hexagon (H) sites with 
a binding energy of 2.1 eV, V and Cr prefers double-bond (D) 
sites with binding
energies of 1.3 and 0.8 eV, respectively. Heavier metals such
as Mn, Fe, and Co do not bond on C$_{60}$. Once the metals
are absorbed on C$_{60}$, each can bind up to 
four hydrogen molecules with
an average binding energy of 0.3-0.5 eV/H$_{2}$. At high 
metal-coverage, we show that a C$_{60}$ can accommodate six D-site
and eight H-site metals, which can reversible 
absorb up to 56 H$_{2}$ molecules, 
corresponding to 7.5 wt\%.
%56 molecularly absorbed
%hydrogens yielding total of 14 transition metals per C$_{60}$
%This corresponds to molecular absorption of 56 H$_{2}$, yielding
%to 7-8 wt\% hydrogen absorption. 
\end{abstract}

\pacs{61.46.+w,68.43.-h,84.60.Ve,81.07.De} \maketitle

An efficient storage media for hydrogen is 
crucial for the advancement of hydrogen and fuel-cell 
technologies\cite{science-review}. There have been a great number of 
reports on the search for new
routes to engineer nanomaterials so that (a) they dissociate H$_{2}$
molecules into H atoms and (b) reversibly adsorb hydrogen molecules
at ambient 
conditions\cite{science-review,chan,tada,miura,gang,dubot,lee,gulseren,gulseren-prl}. 
Much effort has been focused on the
engineering of carbon-based materials such as nanotubes\cite{sefa,litubeprl}
 and metal hydrides such as alanates\cite{hydride}.
 It is found that while hydrogen-carbon interaction is too weak\cite{sefa},
  the metal-hydrogen interaction is too strong for hydrogen
  storage at ambient conditions. 
 Very recently we have shown\cite{tanerprl}
 a novel way to overcome this
 difficulty by forming artificial metal-carbide like structures on
 carbon singled-wall nanotubes (SWNT). From accurate 
first-principles calculations, we show
 that a single Ti-atom adsorbed  on a SWNT can strongly bind up to 
 four hydrogen molecules\cite{tanerprl}. 
 %Remarkably, this adsorption occurs with no energy barrier. 
 At large Ti coverage we find that a (8,0) SWNT can store hydrogen 
 molecules up to 8-wt\%, exceeding the minimum requirement of 
 6-wt\% for practical applications. 
 Even though these results were totally unexpected, we explained them
by a simple  Dewar, Chatt, and 
Duncanson (DCD) model\cite{kubasbook,DCD}, where the interaction is due
to donation of charge from the highest occupied orbital
of the ligand to the metal empty states and a subsequent back
donation from filled d-orbitals into the lowest unoccupied 
orbital of the ligand.

 Here we show that similar
 phenomena also occurs in light-transition metal decorated 
 C$_{60}$ molecules. Below we first discuss several possible 
 absorption sites for a single Ti atom on a C$_{60}$ molecule.
 We then show how a single Ti atom on a C$_{60}$ can bind up to
 four hydrogen molecules via Kubas interaction\cite{kubasbook,DCD}. 
Multiple metal coverage cases, yielding up to 8 wt\% hydrogen absorption,
 are discussed next. Finally, we briefly discuss the results
 for other transition metals from Sc to Co.

\begin{figure}
\includegraphics[scale=0.35,angle=0]{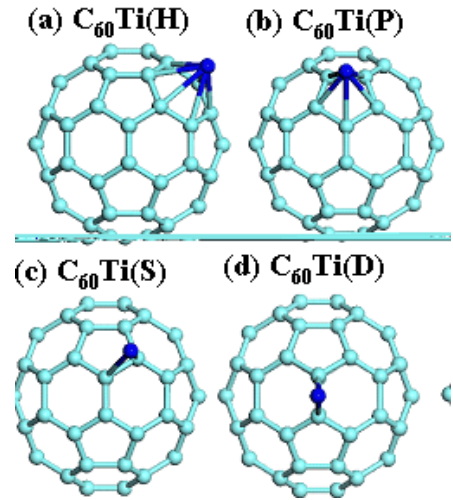}
\caption{
A single Ti atom absorbed at hexagonal (H) (a) and pentagonal (P) (b)
hollow sites, and double (D) and single (S) bond sites of a 
C$_{60}$ molecule, respectively.
%The relevant structural parameters and binding energies of these
%configurations are listed in Table~\ref{table1}.
}
\label{fig1}
\end{figure}

The energy calculations were performed within the plane-wave 
implementation\cite{castep} of the generalized gradient approximation\cite{pbe} 
to DFT. We used Vanderbilt ultra-soft pseudopotentials\cite{usp} treating the 
following electronic states as valence: Ti: $3s^{2}3p^{6}3d^{2}4s^{2}$, 
C: $2s^{2}2p^{2}$ and H: $1s$. 
%TYTYTY
%We carefully tested the convergence of our 
%calculations with respect to the plane-wave cutoff.
The cutoff energy of 350 eV is found to be enough for total 
energies to converge within 0.5 meV/atom.  The calculations are 
carried out in a cubic supercell of 16~\AA\ of side 
for single-metal C$_{60}$ systems and 20~\AA\ 
for full-coverage cases. 
%The system
%is always treated to be metal but only $\Gamma$-point is used since
%we are studying a single molecule in a very large supercell. 
We also carried out spin-polarized calculations for cases where 
the ground state of the metal-coated C$_{60}$ is magnetic. 
Structural relaxations were considered to be converged when forces on
atoms were smaller than 0.2 eV/\AA.

Figure~\ref{fig1} shows the four possible absorption sites on a C$_{60}$ molecule that
are considered in this study. The C$_{60}$ molecule has the truncated-icosahedral
symmetry with twenty hexagons (H), twelve pentagons (P), thirty double C-C bonds (D)
between two hexagons and sixty single C-C bonds (S) which are between pentagon and hexagon
carbon rings. We calculated the binding energy of a single Ti atom at these sites,
which are listed in Table~\ref{table1} along with relevant structural parameters.
The binding energy is defined as 
\begin{equation}
E_{B}(Ti) =E(C_{60})+E_{spol}(Ti) - E_{spol}(C_{60}Ti),
\label{be}
\end{equation}
where $E_{spol}$ is the spin-polarized energy. Hence the positive binding
energy indicates the stability of the system.

\begin{table}
\begin{tabular}{|r||c|c|c|c|} \hline \hline
  &     C$_{60}$Ti(H)  & C$_{60}$Ti(P)  & C$_{60}$Ti(D) & C$_{60}$Ti(S) \\ \hline \hline
  d(Ti-C) (\AA) &    2.27 &  2.33 & 2.10& 2.24  \\  
 d(C-C) (\AA) &   1.42/1.45  & 1.44  & 1.52 & 1.48 \\  
  Q(Ti) (e)  &   1.39 & 1.09 & 0.99 & 0.84\\  
  S(Ti)  &   0.99 $\hbar$ & 1.40 $\hbar$  &  1.13 $\hbar$ &1.54 $\hbar$ \\  \hline
  E$_{B}$(Ti) (eV) &  2.098 &  1.633 & 1.837 & 1.220   \\    \hline \hline
\end{tabular}
\caption{The calculated Ti-C and C-C bond distances, the Mulliken charges, spins and
the binding energies for a single Ti atom absorbed at four different 
sites of a C$_{60}$ molecule as shown in Fig.\ref{fig1}.
For bare C$_{60}$ the calculated double and 
single bond lengths  are 1.44 \AA $\;$ and 1.38 \AA, respectively.}
\label{table1}
\end{table}

The results in  Table~\ref{table1} indicate that H-site is the 
most stable configuration for C$_{60}$Ti with a binding energy of 2.1 eV.
The D-site comes next with the shortest Ti-C bonds among the four
possible configurations. For all cases, we have about one electron
charge transfer to the C$_{60}$ molecule. The S-site is the
least stable absorption site and therefore we do not consider it
any further. 

\begin{figure}
\includegraphics[scale=0.60,angle=0]{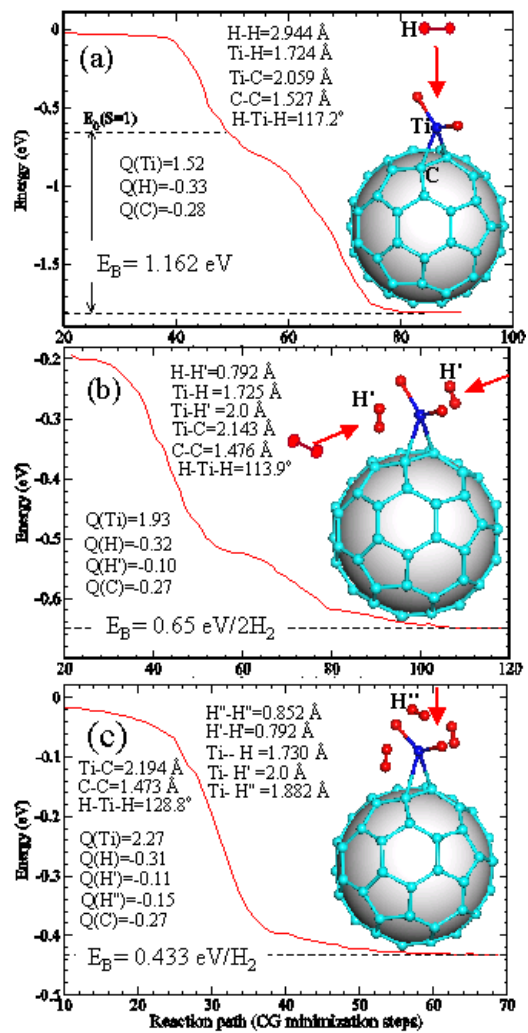}
\caption{
Energy versus reaction paths for successive 
dissociative and molecular adsorption of 
H$_{2}$ over a single C$_{60}$Ti(D). 
(a) H$_{2}$ + C$_{60}$Ti(D) $\rightarrow $C$_{60}$Ti(D)H$_{2} $. 
(b) 2H$_{2}$+  C$_{60}$Ti(D)H$_{2} $ $\rightarrow$ C$_{60}$Ti(D)H$_{2}$-2H$_{2}$. 
(c) H$_{2}$+ C$_{60}$Ti(D)H$_{2}$-2H$_{2}$ $\rightarrow $ C$_{60}$Ti(D)H$_{2}$-3H$_{2}$. 
The zero of energy is taken as the sum of the energies of two reactants. 
}
\label{fig2}
\end{figure}

Figure 2(a) shows the energy variation from {\it spin-unpolarized}
calculations as a single H$_2$ molecule approaches C$_{60}$Ti(D).
The energy first decreases slowly as the hydrogen gets closer to the C$_{60}$
and Ti. However, as the charge overlap gets large, the H$_{2}$
 molecule is attracted 
towards the Ti atom with a sudden decrease in the energy. At this point, 
the H$_{2}$  molecule is still intact with a significantly increased H-H bond 
length of 0.9 \AA. The second sudden decrease in energy is achieved by 
dissociating the H$_{2}$ molecule into two H atoms. At this point, the H-H 
distance increases from 0.9 \AA \; to 2.94 \AA. 
The interaction between H$_{2}$ and C$_{60}$Ti(D) is always attractive 
and therefore H$_{2}$ is  absorbed onto a Ti atom 
without any energy barrier. The final geometry is shown in the inset to 
Fig.~2(a), with relevant structural parameters given\cite{animations}. 
In order to calculate the binding energy for this dissociative adsorption,
we calculate total energies of the initial C$_{60}$Ti(D)  and H$_{2}$
states and the final  C$_{60}$Ti(D)H$_{2}$ state (dashed lines in Fig.~1(a))
from spin-polarized calculations. We obtained the binding energy to
be 1.16 eV (Fig.~2(a)).

Figure~2(b)  shows the energy  variation as two hydrogen molecules 
approach the Ti atom; one from each  side of the TiH$_{2}$  group. 
As in the case of single adsorption, the energy 
always decreases, first slowly and later very rapidly at which point 
both hydrogen molecules are strongly attached to the C$_{60}$Ti(D)H$_{2}$ system. 
In the final configuration, we note that two H$_{2}$ molecules were 
rotated by 90$^{\rm{o}}$ as shown in Fig.~2(b). 
We denote the final product as  C$_{60}$Ti(D)H$_{2}$-2H$_{2}$.
The total energy change upon adsorption is 
about 0.65 eV (i.e. 0.325 eV/H$_{2}$). 
Unlike the first adsorption, the two hydrogen molecules are in intact but 
with a rather elongated bond length of 0.79 \AA. 

Figure~2(c) shows the energy evolution when a fourth hydrogen molecule 
approaches the  C$_{60}$Ti(D)H$_{2}$-2H$_{2}$   system from the top. 
The energy again 
decreases continuously, indicating a zero-energy barrier. 
The final product, denoted as C$_{60}$Ti(D)H$_{2}$-3H$_{2}$, 
is shown in the inset 
with the relevant structural parameters. The energy gained by 
the fourth adsorption, 0.433 eV/H$_{2}$, is slightly larger than for 
the previous one. The H-H distance of  the top H$_{2}$ is 0.85 \AA.  
Several attempts to add a fifth hydrogen 
molecule at a variety of positions failed, suggesting 
a limit of 4 H$_{2}$/Ti. However in the view of a recent study\cite{mh12}
which shows that it is possible to attach twelve hydrogen atoms to a single
transition metal, it is quite possible that there may be other transition
paths that could yield  more than four hydrogen molecules per C$_{60}$Ti. 
%Finally, we also consider other isomers of C$_{60}$Ti(D)$_{2}$-3H$_{2}$
%where all the hydrogen molecules are absorbed molecularly. In the case of
%Ti-decorated SWNT, four molecular absorption isomer was found to be the
%ground state. However, for the C$_{60}$Ti(D)$_{2}$-3H$_{2}$ system,
%such isomer has about 0.2 eV higher energy than one shown in Fig.~2(c).

%
\begin{figure}
\includegraphics[scale=0.50,angle=0]{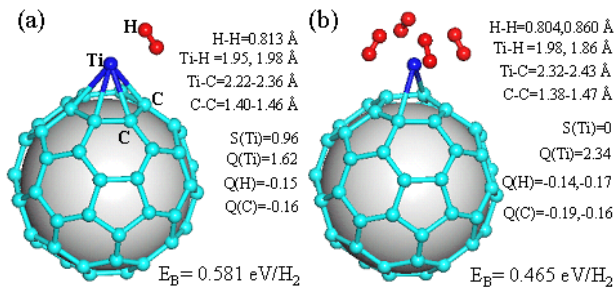}
\caption{
Molecular absorption of a single (a) and four H$_{2}$ on
C$_{60}$Ti(H). The binding energies and relevant structural
parameters are also given.
}
\label{fig3}
\end{figure}

We next discuss the hydrogen absorption properties of a Ti 
on the H-site of a C$_{60}$ as shown in Fig.~3. For the first 
hydrogen absorption, we find that the H$_{2}$ stay intact unlike the
case of Ti  on a D-site where the first H$_2$ is absorbed
dissociatively as shown in Fig.~2(a). This is due the fact that
Ti is absorbed on H-site very strongly and therefore there is
not enough charge left on Ti to transfer to the $\sigma^{*}$ of H$_{2}$
to dissociate it\cite{tanerprl}. The binding energy is
about 0.58 eV and H-H bond length is 0.813 \AA. Introducing 
additional hydrogen molecules yield more molecular absorption
without any energy barrier up to 4H$_{2}$/Ti, which is 
denoted as C$_{60}$Ti(H)-4H$_{2}$ and shown in Fig.~3(b). 
The final optimized structure is very symmetric, all 
hydrogen molecules benefit equally from the bonding with
Ti atom. The average binding energy per H$_{2}$ is about
0.465 eV, slightly smaller than the one of the first
absorption. We have also calculated the binding energy
for the isomer,
C$_{60}$Ti(H)H$_{2}$-3H$_{2}$, where the first H$_2$ is
bonded to Ti dissociatively. We find that this isomer
is about 0.1 eV higher in energy than 
C$_{60}$Ti(H)-4H$_{2}$.

\begin{figure}
\includegraphics[scale=0.45,angle=0]{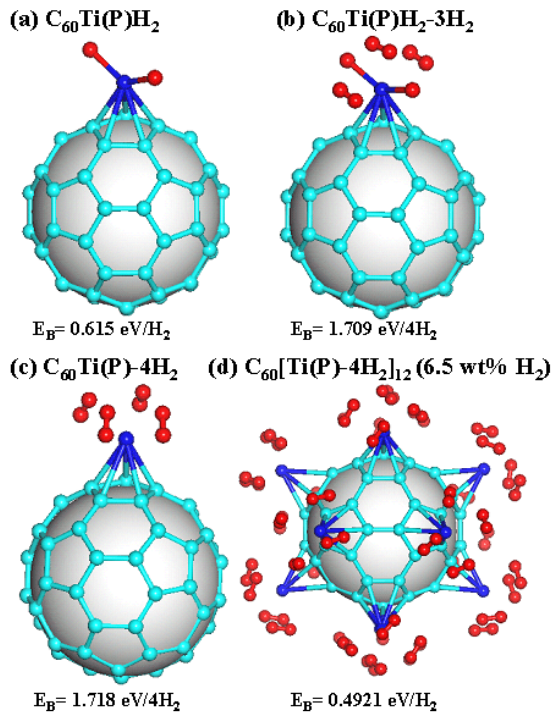}
\caption{
The optimized structure of one (a) and two isomers (b-c) of 
four hydrogen molecules absorbed on a single Ti atom on P-site of a 
C$_{60}$. (d) Full coverage case where all twelve pentagon
sites are decorated by Ti, 
each of which absorbs four H$_{2}$ molecules.
}
\label{fig4p}
\end{figure}

The results for the absorption of hydrogens on 
the P-site Ti are summarized in Fig.4. The first absorption
is found to be dissociative without activation energy. The
binding energy is about 0.615 eV, significantly smaller than
that of Ti on D-site (see Fig.~2(a)), but the absorption is
still dissociative unlike Ti on H-site (see Fig.3(a)). 
The two isomers, C$_{60}$Ti(P)H$_{2}$-3H$_{2}$  (Fig.~4(b)) and 
C$_{60}$Ti(P)-4H$_{2}$ (Fig.~4(c)), are found to be 
almost degenerate and yield a binding energy of 0.43 eV/H$_{2}$.

To this point we have discussed the interaction of H$_{2}$ with a single Ti atom 
bonded to a C$_{60}$, but clearly one can imagine attaching more Ti to a C$_{60}$, 
thereby increasing the hydrogen storage capacity. In order to show the feasibility 
 of this approach, we consider several cases. Fig.~4(d) shows the
 full coverage case where all the P-sites of a C$_{60}$ molecule is absorbed
 by Ti atoms, each of which binds four H$_{2}$ molecules. The calculated 
average binding  energy  for each hydrogen molecule is 0.492 eV/H$_{2}$.
This is slightly higher than the single-coverage case  as shown
in Fig.~4(c). This is  due to the more symmetric configuration of the 
full coverage case, where C$_{60}$ does not have to distort locally
as in single-coverage case. In fact, we see the same effect in the binding energy
of Ti atom on P-site, which is   1.633 eV (see Table~\ref{table1}),
for a single  Ti atom. However it becomes
2.115 eV/Ti when all the P-sites are absorbed by Ti. These results
are quite promising, suggesting that the full coverage
 system is more stable 
than the single-atom coverage.

\begin{table*}
\begin{tabular}{|r||c|c|c|c|c|c|c|} \hline \hline
 Properties &     C$_{60}$Sc(H)  & C$_{60}$Ti(H)  & C$_{60}$V(D) & C$_{60}$Cr(D) &
    C$_{60}$Mn(D)  & C$_{60}$Fe(D)  & C$_{60}$Co(D)    \\ \hline  
  d(TM-C) (\AA) &    2.29 &  2.27 & 2.18& 2.09  & 2.30  & 2.28 & 2.26 \\  
 d(C-C) (\AA) &   1.42/1.46 & 1.42/1.45  & 1.46 & 1.50 & 1.45 &  1.45 & 1.45 \\  
  Q(TM) (e)  &  1.55 & 1.39 & 0.78 & 0.91  &  0.97 &   0.86&  0.76 \\  
  S(TM)  &   0.24 $\hbar$ & 0.99 $\hbar$  &  2.0 $\hbar$ & 2.35 $\hbar$  
  & 2.99 $\hbar$  &2.53 $\hbar$  & 1.54 $\hbar$ \\  \hline
  E$_{B}$(TM) (eV)     & 2.127  & 2.098  & 1.308  & 0.760  &-0.017   & -0.130 & -0.503\\    
E$_{B}$(H$_{2}$) (eV) &  0.300 &  0.454 &  0.497 &  0.239 &  -0.092 &   -    & -     \\ \hline  \hline
\end{tabular}
\caption{The calculated TM-C and C-C bond distances, the Mulliken charges, spins and
the binding energies for a single TM  atom (TM=Sc, Ti, V, etc)
absorbed on a C$_{60}$ molecule. 
}
\label{table2}
\end{table*}

Figure~5(a) shows another case where 
six Ti atoms absorbed on the D-sites (those sites along the two-fold axis
of the C$_{60}$ molecule). As in the case of full pentagonal-coverage,
the average binding energy per H$_{2}$ is slightly increased to
0.592 eV from the single Ti value of 0.559 eV. 
Interestingly, one can further add eight more Ti-atoms at  the 
hexagonal faces of the C$_{60}$ molecules (those along the [111]
directions), yielding total 14 Ti atoms per C$_{60}$. The full
hydrogen absorbed case of this coverage is shown in Fig.~5(b). The
average binding energy is 0.522 eV/H$_{2}$, in excellent agreement
with the 0.505 eV/H$_{2}$ based on the single Ti-atom cases.
This indicates that the 14 Ti atoms and 56 hydrogen
molecules shown in Fig.~4(b) are not too close to each other
and therefore 
suggest that the system has the capacity to have many Ti and hydrogen. 
 In fact, the configuration shown in Fig.~5(b), which have the chemical formula,
 C$_{60}$Ti$_{14}$ H$_{56}$,  stores  approximately 
 7.5-wt\% hydrogen.
 %assuming all H$_{2}$ can be retreated. 
 %If we assume that we can get out only those H$_{2}$ which 
 %are molecularly bonded, then the estimated capacity is
 %6.7 wt\%, still exceeding the minimum requirement of 6 wt\%
 %for practical applications.

%
\begin{figure}
\includegraphics[scale=0.50,angle=0]{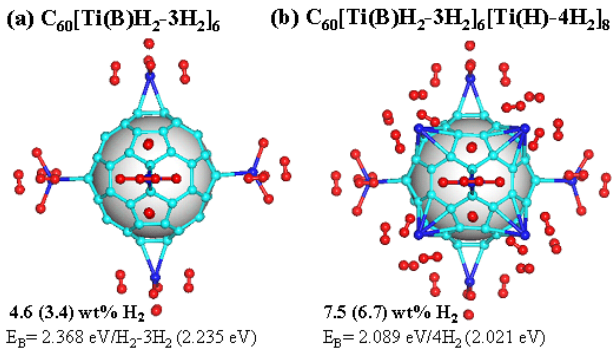}
\caption{
Two high-density  coverage on Ti-coated C$_{60}$, 
corresponding to 4.3 wt\% and 7.5 wt\% hydrogen storage
when  all hydrogens can be released and 
3.4 wt\% and 6.7\% storage when only molecularly
absorbed hydrogens are released.
}
\label{fig4}
\end{figure}

It is important to know if the results reported above 
for the C$_{60}$Ti system   hold for other transition metals.
Therefore we have also studied the transition metals
from Sc to Co. Table~\ref{table2} summarizes our results
for the binding energies and relevant structural
parameters. 
Briefly we find that while Sc and Ti prefers the 
H-site, V and Cr prefers the D-sites. The binding
energies monotonically decreases as we move from
left to the right of the periodic table. In fact
for heavy transition metals such as Mn, Fe, and Co,
we obtained negative binding energy, indicating that
the system is not stable against C$_{60}$ and isolated
metal atom formation. We also confirmed that similar
hydrogen absorption happens for the other transition
metals. The last line in Table~\ref{table2} indicates
the average binding energy per H$_{2}$ in C$_{60}$TMH$_{2}$-3H$_{2}$
configurations where TM=Sc, V, etc.  It increases from
0.3 eV/H$_{2}$ for Sc to 0.5 eV/H$_{2}$ for V and then
decreases to 0.239 eV/H$_{2}$ for Cr and then the system
becomes unstable for heavier transition metals. 

In conclusion,  using the state-of-the-art  first-principles calculations
we show that light-transition metal-decorated C$_{60}$ molecules 
exhibit remarkable hydrogen storage properties. Metals bonded on the 
D-site dissociate the first H$_{2}$ molecule without any activation 
barrier and then reversible and molecularly bond three more hydrogens. 
However metals absorbed at the H-sites binds four hydrogen molecules
without dissociating it. Combining these two binding sites we
shows that a single C$_{60}$ molecule is able to bind up to
56 H$_2$ molecules, corresponding to 7.5 wt\% storage capacity.
These results combined with our previous work in Ti-decorated
SWNT\cite{tanerprl} suggest a new direction to high-capacity hydrogen
storage materials by decorating light-transition metals on 
nanostructured materials. The nature of hydrogen bonding is 
explained by DCD model\cite{kubasbook,DCD}
and has the right strength for room temperature 
reversible hydrogen storage.

This work was partially supported by DOE under Grant 
No.  DEFC36-04-GO14280.
% and by NSF under Grant No. INT0115021.  

\end{document}